\documentclass[10pt,aps,prb,twocolumn, nofootinbib]{revtex4-1}

\usepackage[caption=false]{subfig}

\usepackage{amssymb,amsmath}
\usepackage{subfig}
\usepackage{graphicx}
\usepackage{color}
\usepackage{hyperref}
\usepackage{listings}
\usepackage{my_math}
\hypersetup{
    bookmarks=true,         
    unicode=false,          
    pdftoolbar=true,        
    pdfmenubar=true,        
    pdffitwindow=false,     
    pdfstartview={FitH},    
    pdfauthor={Kevin Multani},     
    colorlinks=true,       
    linkcolor=blue,          
    citecolor=black,        
}

\graphicspath{{figures/}}

\begin{document}

\definecolor{dkgreen}{rgb}{0,0.6,0}
\definecolor{gray}{rgb}{0.5,0.5,0.5}
\definecolor{mauve}{rgb}{0.58,0,0.82}

\def\bibsection{\section*{\refname}}

\lstset{frame=tb,
  	language=Matlab,
  	aboveskip=3mm,
  	belowskip=3mm,
  	showstringspaces=false,
  	columns=flexible,
  	basicstyle={\small\ttfamily},
  	numbers=none,
  	numberstyle=\tiny\color{gray},
 	keywordstyle=\color{blue},
	commentstyle=\color{dkgreen},
  	stringstyle=\color{mauve},
  	breaklines=true,
  	breakatwhitespace=true
  	tabsize=3
}

\title{Role of Pore Dilation in Molecular Transport through the Nuclear Pore Complex: Insights from Polymer Scaling Theory}
\author{Atsushi Matsuda}
\author{Mohammad R. K. Mofrad}
\email[]{mofrad@berkeley.edu}
\affiliation{Molecular Cell Biomechanics Laboratory, Departments of Bioengineering and Mechanical Engineering, University of California Berkeley, Berkeley, California 94720, USA}
\date{\today}

\begin{abstract}
\noindent
\textbf{Abstract}: 
Recent studies have suggested that the Nuclear Pore Complex (NPC) plays a significant role in mechanotransduction. When a force is exerted, the NPC's diameter widens, leading to an increased molecular flux into the nucleus. 
In this study, we sought to further explore this phenomenon and quantitativelly assess the impact of pore dilation on molecular transport through the NPC.
Utilizing the scaling theory of polymers, we developed a theoretical model to examine the relationship between pore size and the molecular transport rate. Our model posits that the mesh structure inside the pore, formed by FG-Nups, significantly influences the transport rate. Consequently, we propose that the transport rate is exponentially related to the pore size.
To validate our model, we conducted extensive Brownian dynamics simulations. Our results demonstrated that the model accurately represents the transport dynamics except for exceptionally small molecules. For these molecules, the local mesh structure becomes less significant, and instead, they perceive the global structure of the pore.
We also identified a critical threshold value, which allows for an estimation of whether a given molecule falls within the scope of our model. Our findings provide valuable insights into the dynamics of molecular transport in the NPC and pave the way for future research on the NPC's role in mechanotransduction.

\noindent
\textbf{Significance}: 
The Nuclear Pore Complex (NPC) is a critical component of the cell, acting as a selective gateway for molecular traffic in and out of the nucleus. Understanding the mechanisms underpinning its filtration abilities is not only crucial for a deeper knowledge of cellular function, but also has potential applications in bioengineering. In this study, we suggest that a mesh-like structure of polymers, housed within the NPC, plays a significant role in these processes. Drawing on polymer scaling theory, we have developed a mathematical formula that elucidates the relationship between the transport rate of molecules and the NPC's structure. 
Our study proposes a quantitative understanding of NPC function, demonstrating how structural changes within the NPC significantly alter the behavior of molecular transport. 

\noindent
\textbf{Classification}: Biological Sciences (Biophysics and Computational Biology)

\end{abstract}
\maketitle

\begin{figure}[b]
\centering
\includegraphics[width=0.48\textwidth]{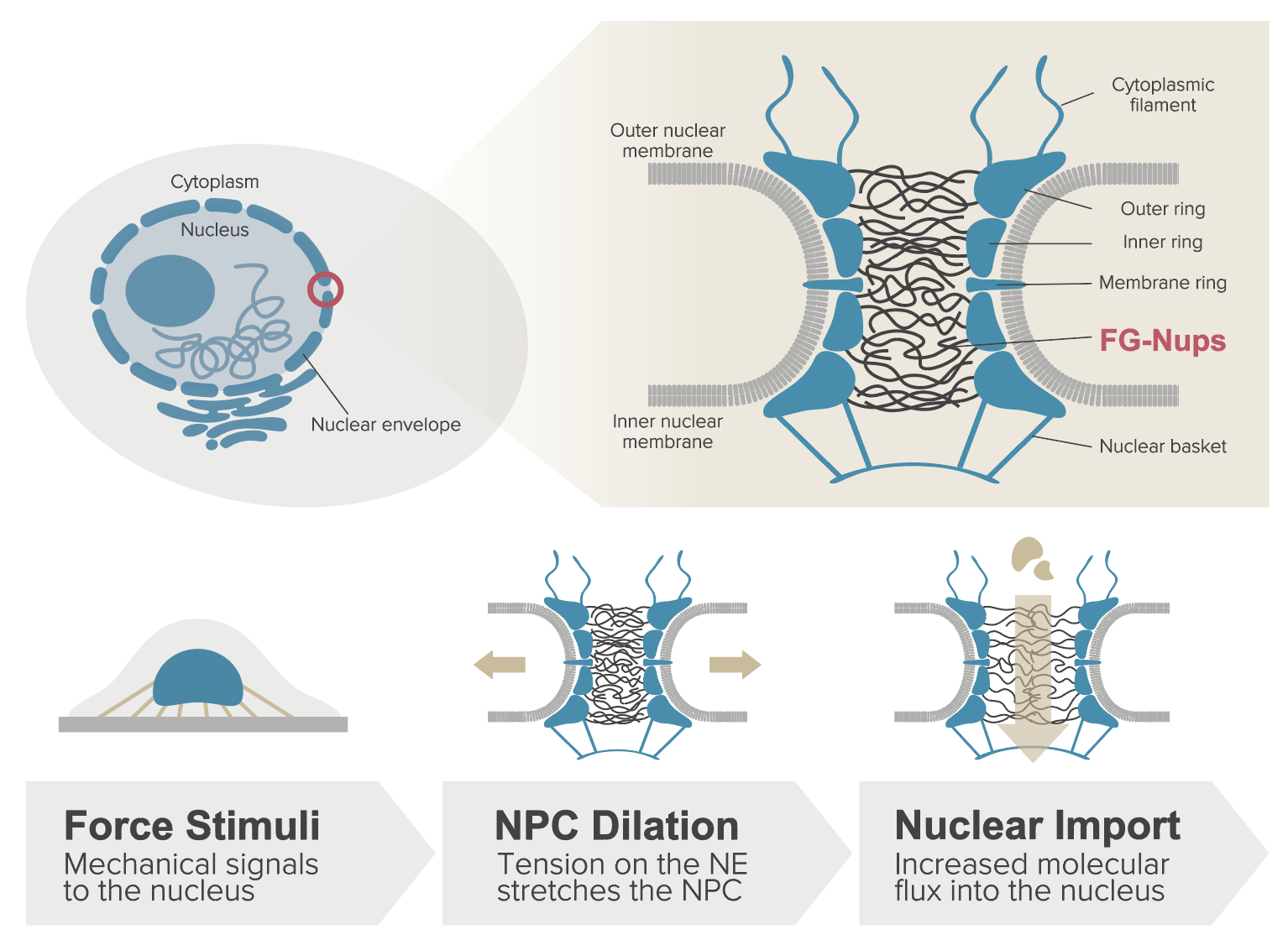}
\caption{The nuclear pore complex (NPC) and its potential involvement in mechanotransduction. Top panel depicts the location and the intricate structure of the NPC. Bottom panel illustrates the hypothesized role of the NPC in the mechanotransduction process.}
\label{fig:intro}
\end{figure}

\section{Introduction\label{sec:intro}}
The nuclear pore complex (NPC) is a nanopore located in the nuclear membrane, serving as a spatial conduit between the cytoplasm and the nucleus \cite{hoogenboom2021physics,jamali2011nuclear,cowburn2023improving}. 
Acting as the exclusive passageway between these two compartments, the NPC regulates the traffic of all molecules entering and exiting the nucleus.
The NPC is comprised of 500-1000 proteins known as nucleoporins, collectively forming an hourglass-shaped channel with a diameter of 40-60 nm \cite{huang2023structure,bley2022architecture,petrovic2022architecture,kim2018integrative}. 
Attached to the inner wall of this channel are intrinsically disordered proteins, termed FG-Nups, which occupy the interior space of the channel \cite{peyro2021fg,peyro2021nucleoporins}.
These FG-Nups feature numerous phenylalanine and glycine motifs, referred to as FG-motifs. 
The FG-motifs are capable of forming transient bonds with each other via hydrophobic interaction \cite{yu2023visualizing,dekker2023phase,aramburu2017floppy}.
The hydrophobic interactions among FG-motifs facilitate the formation of a dynamic polymer mesh structure within the channel of the NPC.
This strucutre inherently restricts the size of molecules that can transverse the NPC. 
However, if a molecule is capable of forming hydrophobic interactions with FG-motifs, this size limitation can be expanded, allowing larger molecules to pass through \cite{matsuda2021free,zheng2023self,winogradoff2022percolation}.

An emerging hypothesis has recently suggested a potential involvement of the NPC in mechanotransduction \cite{matsuda2022nuclear,kalukula2022mechanics,donnaloja2019mechanosensing}.
Mechanotransduction refers to the process in which cells transform mechanical stimuli into biochemical signals, enabling them to sense and respond to their physical environment \cite{mofrad2014cellular,jahed2014mechanotransduction}.
This process encompasses several molecular pathways, each facilitating the transformation of mechanical stimuli into biochemical signals. 
Recent research suggests that the NPC may play a significant role in one of these pathways, as detailed below: 
(1) Applied mechanical force deforms the cell's nucleus and puts tension on the nuclear envelope; (2) This tensile force stretches the NPC, leading to an increase in pore diameter; (3) The expansion of the NPC facilitates an elevated molecular flux into the nucleus, potentially influencing gene transcription.

To explore the above hypothesis, several research groups have examined the changes in the pore size of the NPC in response to mechanical stimuli. 
Elosegui-Artola et al. \cite{elosegui2017force,elosegui2018control,andreu2022mechanical} applied tension to the nuclear envelope by either enhancing substrate stiffness or exerting force via atomic force microscopy. 
They observed pore dilation accompanied by an increased molecular influx into the nucleus. 
In contrast, Zimmerli et al. \cite{zimmerli2021nuclear} reduced tension on the nuclear envelope through energy depletion or the application of hyperosmotic shock, resulting in the observation of NPC constriction.
Additionally, there are several hypotheses regarding the molecular composition that may facilitate the dilation and constriction of the NPC \cite{mosalaganti2022ai,li2022near,solmaz2011molecular}.

While more evidence emerges about the structural flexibility of the NPC, our understanding of how pore dilation affects molecular flux remains limited.
Recently, Klughammer et al. \cite{klughammer2023diameter} examined the relation between pore size and molecular flux,
employing artificial NPC mimics for their investigation. 
They proposed a quadratic function to express this relationship, denoted as:
\begin{equation}
    k_{AB} \sim (D-D_0-d)^2.
\end{equation}
Here, $k_{AB}$ is the transport rate, $D$ is the pore diameter, $D_0$ is the shifting factor corresponding to the height of tethered polymers, and $d$ is the cargo diameter. 
This formula assumes that the transport rate is proportional to the effective cross-sectional area of the ``void'' space accessible to the transported molecules.
While this formula holds true for pore sizes larger than $(D_0+d)$, the transport rate when the pore diameter falls below this threshold remains ambiguous. 
Notably, the typical fluctuation in NPC diameter is within the 40-60 nm range \cite{matsuda2022nuclear}, 
which falls beneath the aforementioned critical diameter.

In this study, we have developed a model formula to estimate the transport rate when the pore diameter falls below the critical threshold.
In this particular regime, there exists no ``void'' space within the pore as FG-Nups occupy the entirety of the interior space.
In this environment, the molecules undergoing transport are immersed within the polymer solution of FG-Nups,
and their motion is constrained by the mesh size of the polymer solution.
To account for this situation, we incorporated the principles of polymer scaling theory \cite{rubinstein2003polymer,de1979scaling,doi1988theory} into our model. 
This approach resulted in deriving an exponential relationship between the transport rate and the pore diameter.

The remainder of this paper is structured as follows. 
In the Model section (\ref{sec:model}), we derive the relation between the transport rate and the pore diameter using scaling theory of polymers. 
The Simulation section (\ref{sec:sim}) details the simulation model that we employed to validate the derived model formula. 
In the Results section (\ref{sec:res}), we present our simulation results and compare them with our theoretical model. 
In the Discussion section (\ref{sec:disc}), we discussed the limiation of our model. 
Lastly, the Conclusion section (\ref{sec:conc}) summarizes our findings. 

\section{Model\label{sec:model}}
In this section, we derive the model formula that describes the transport rate in terms of the pore diameter.
We consider the system where $n_\mathrm{FG}$ FG-Nups are anchored to the inner surface of the NPC channel. 
Each FG-Nup is conceptualized as polymers containing $N$ monomers, following the scaling law, $R \approx b N^\nu$,
where $R$ is the radius of gyration, $b$ is the Kuhn length, and $\nu$ is the Flory exponent. 

We focus on the scenario where the polymer concentration within the NPC exceeds the overlap concentration, a situation typically referred to as a semi-dilute polymer solution.
In a semi-dilute polymer solution, the mesh-like structure of the polymer solution is characterized by a length scale known as the correlation length, $\xi$. 
Conceptually, the correlation length can be interpreted as the average distance between two polymers, thereby serving as a representative measure of the mesh size of the structure. 
For the remainder of this paper, we will refer to the correlation length as the ``mesh size".
On the basis of the scaling law, the mesh size can be quantified using the volume fraction of polymers, $\phi$, as follows \cite{rubinstein2003polymer,de1979scaling}:
\begin{equation}
\label{eq:xi_ori}
\xi \approx b \phi^{-\nu/(3\nu-1)}.
\end{equation}
We suppose that the volume fraction can be expressed as $\phi \approx AN/D^2$.
In this equation, $D$ is the diameter of the pore and $A = 4v_\mathrm{mon} n_\mathrm{FG} / \pi h$ is the prefactor having a unit of squared area, where $v_\mathrm{mon}$ is the monomer volume and $h$ is the height of the pore.

The overlap concentration can be defined as $\phi^* \approx Nb^3/R^3 \approx N^{1-3\nu}$ \cite{rubinstein2003polymer,de1979scaling}. 
This allows the pore diameter and the mesh size at the overlap concentration to be denoted as $D^* \approx A^{1/2} N^{3\nu/2}$ and $\xi^* \approx bN^\nu$, respectively.
With these scale parameters, we can derive the normalized form of equation \ref{eq:xi_ori} as:
\begin{equation}
\label{eq:xi}
\xi/\xi^* \approx (D/D^*)^{2\nu/(3\nu-1)}.
\end{equation}

\begin{figure}[t]
\centering
\includegraphics[width=0.35\textwidth]{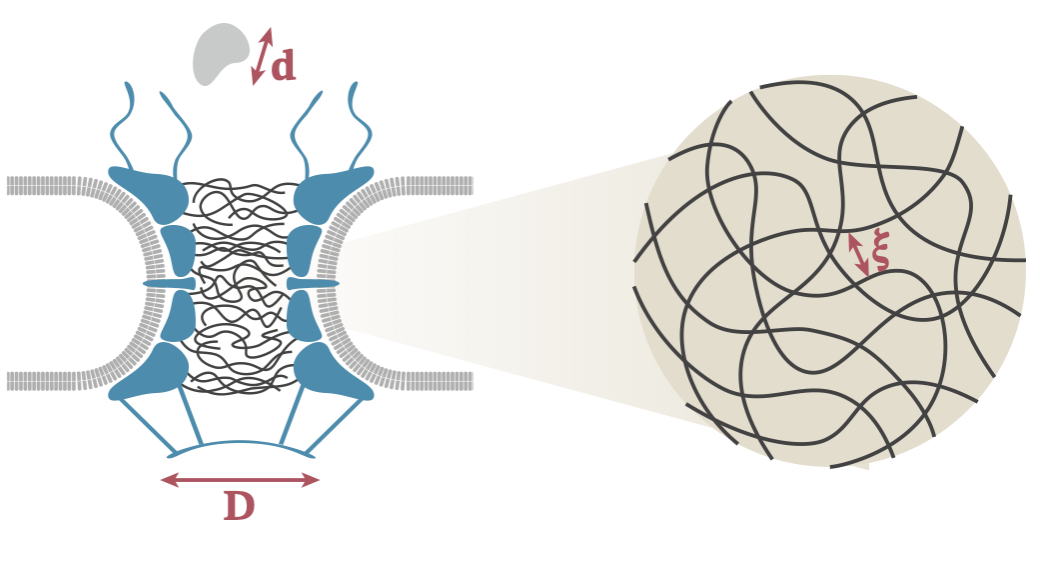}
\caption{Key scales of interest in model formulation. $D$: diameter of the pore, $d$: diameter of the cargo, $\xi$: mesh size of the polymer solution.}
\label{fig:model}
\end{figure}

Next, we estimate the free energy, $\Delta G$, required to insert a molecule with diameter $d$ into the polymer solution. 
There are three length scales of interest in the set-up: the molecule's diameter $d$, the pore's diameter $D$, and the mesh size $\xi$ (Fig. \ref{fig:model}). 
While a molecule is being inserted into the channel, it interacts with the local structure of the polymer mesh, not the entirety of the pore's geometry. 
Therefore, we posit that the relevant length scales for the free energy calculation are solely the molecular size $d$ and the mesh size $\xi$, 
resulting $\Delta G / \kBT \approx h(d/\xi)$. 
Here, $\kB$ is Boltzmann constant, $T$ is temperature, and $h(\cdot)$ is a dimensionless function. 
Additionally, we assume that the free energy is directly proportional to the polymer volume fraction, $\phi$ \cite{odijk1996protein}. 
This is underpinned by the rationale that the number of monomers displaced by the molecule's insertion scales with $\phi$, and displacing each monomer requires a specific amount of work.
By incorporating Eq. \ref{eq:xi_ori} into this consideration, we get
\begin{equation}
\label{eq:free}
\frac{\Delta G}{\kBT} \approx \gamma \left( \frac{d}{\xi}\right)^{(3\nu-1)/\nu},
\end{equation}
where $\gamma$ is a dimensionless constant. 

Lastly, we calculate the transport rate of the molecule through the pore, denoted as $k_\mathrm{AB}$. 
The Arrhenius equation tells us that the transport rate changes exponentially with the activation energy \cite{hanggi1990reaction}. 
We approximate the activation energy as $\Delta G$, under the assumption that the most significant energy barrier during the transport process is the insertion of the molecule into the pore.
Thus, we have:
\begin{equation}
\label{eq:rate_ori}
k_\mathrm{AB} = k_0 \exp \left( -\frac{\Delta G}{\kBT} \right),
\end{equation}
where $k_0$ is the transport rate of the molecule with an infinitesimally small diameter.
Combining Eq. \ref{eq:xi}, \ref{eq:free}, and \ref{eq:rate_ori}, we derive:
\begin{equation}
\label{eq:k_AB}
\tilde{k}_\mathrm{AB}
= \exp \left( -\gamma \tilde{d}^\alpha \tilde{D}^{-2} \right).
\end{equation}
Here, $\tilde{k}_\mathrm{AB} = k_\mathrm{AB}/k_0$, is the normalized transport rate while $\tilde{d} = d/\xi^*$ and $\tilde{D} = D/D^*$ are the normalize diameter of the molecule and the pore, respectively. $\alpha = (3\nu-1)/\nu$ is a parameter that depends on the polymer characteristics.

\section{Simulation\label{sec:sim}}
We will explore the validity of our proposed model (as per Eq. \ref{eq:k_AB}) through Brownian dynamics simulations \cite{moussavi2016rapid,moussavi2011biophysical,moussavi2011brownian}. 
In this section, we present the details of our simulation methodology.

\begin{figure}[t]
\centering
\includegraphics[width=0.48\textwidth]{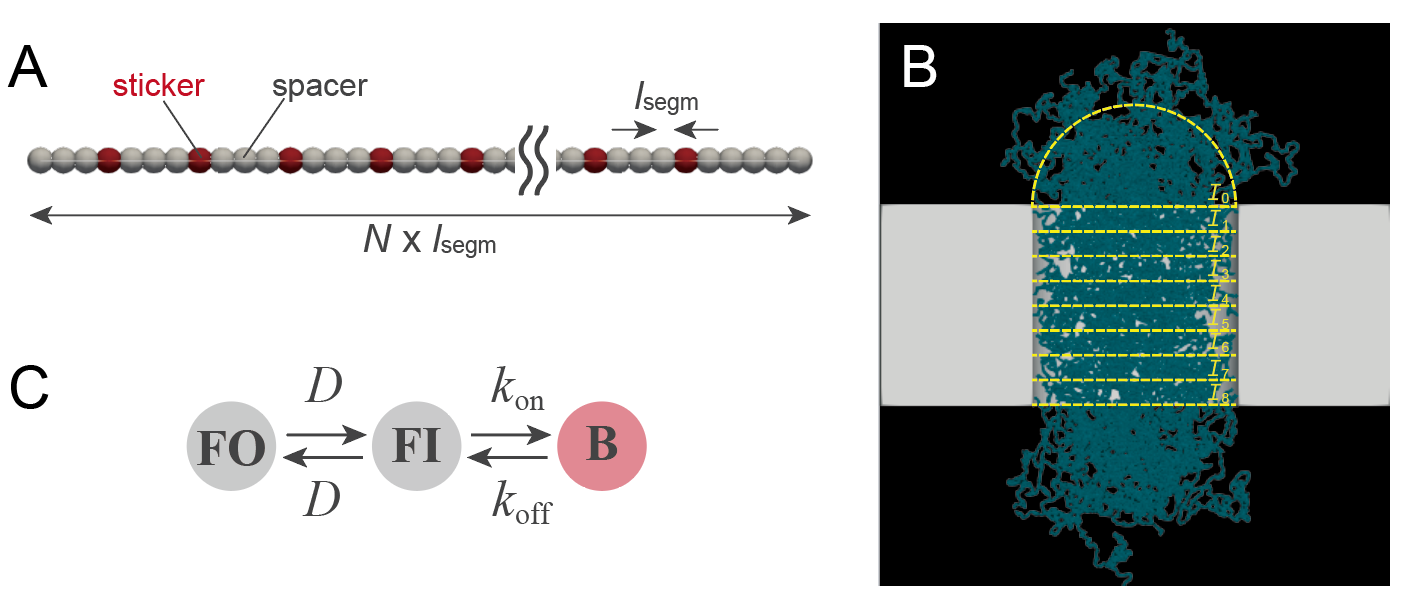}
\caption{Simulation Setup. (A) FG-Nup represented by the spacer-sticker model with a sticker placed every four spacers. (B) Representation of the central channel of the NPC modeled as a cylinder with FG-Nups tethered to the inner wall of the channel. Yellow dashed lines indicate boundaries of each interface for the Forward Flux Sampling (FFS) simulation.
(C) Transition dynamics of the sticker beads among free outside (FO), free inside (FI), and bound (B) states. Transitions between FO and FI are dictated by diffusion, while transitions between FI and B follow a Monte Carlo approach, dictated by the binding and unbinding rates, $k_\mathrm{on}$ and $k_\mathrm{off}$, respectively.}
\label{fig:sim}
\end{figure}

\subsection*{Coarse-grained model}
The simulations were conducted in a $L_x \times L_y \times L_z$ simulation box with periodic boundary conditions applied on each face. To model the nuclear envelope (NE), a partition of thickness $h$ was placed at the half plane of the simulation box. A cylindrical hole of diameter $D_\mathrm{pore}$ was placed in the center of the NE to represent the nuclear pore complex (NPC). The NE and inner wall of the NPC were treated as hard surfaces.

We modeled the inner wall of the NPC by attaching $n_\mathrm{FG}$ pieces of FG-Nups, each of which was represented as a series of bead-springs containing $N$ beads, with each bead having a diameter of $l_\mathrm{segm}$ (Fig. \ref{fig:sim}A). 
The beads were classified into two types: stickers, representing FG-motifs, and spacers, representing other residues \cite{ginell2022introduction}. 
To reflect the ratio of FG-motifs in FG-Nups of the yeast NPC, we placed one sticker in every three spacers \cite{yamada2010bimodal,fragasso2021designer}. 
To model the transported molecules, we used a spherical cargo of diameter $d_\mathrm{cargo}$, whose surface was discretized into a set of beads of the diameter $l_\mathrm{segm}$. 
The cargo surface beads were classified as stickers or spacers based on whether they had an attraction with FG-repeats or not.

\subsection*{Brownian dynamics}
We simulated the dynamics of the system by the overdammped Langevin equation \cite{reif1967fundamentals,underhill2004coarse}, 
\begin{equation}
\label{eq:langevin}
    \zeta \dv{\rr_i}{t} = - \pdv{U}{\xx_i} + \boldsymbol{\Lambda}_i,
\end{equation}
where $\rr_i$ is the position of bead $i$, $\zeta$ is the friction coefficient, $U$ is the potential energy function, and $\boldsymbol{\Lambda}_i$ is the random force on bead $i$. 
The random force satisfies, 
\begin{gather}
    \expval{\Lambda_{i\alpha}(t)} = 0 \\
    \expval{\Lambda_{i\alpha}(t) \Lambda_{j\beta}(t')}  = 2\zeta \kBT \delta(t-t')\delta_{ij}\delta_{\alpha\beta},
\end{gather}
where $\kB$ is Boltzmann constant, $T$ is temperature, $\delta(t-t')$ is Dirac delta function, and $\delta_{ij}$ and $\delta_{\alpha\beta}$ are Kronecker's delta. 
Subscripts $\alpha$ and $\beta$ indicate $\alpha$- and $\beta$- component of the vector.
The potential energy, $U$, is written as,
\begin{equation}
    U = 
    \sum_{\expval{i,j}} U^{\mathrm{bond}}_{i,j} + 
    \sum_{\expval{i,j,k}} U^\mathrm{angle}_{i,j,k} + 
    \sum_{i,j} U^\mathrm{steric}_{i,j},
\end{equation}
where $U^\mathrm{bond}_{i,j}$ is the elastic energy of FG-Nups, $U^\mathrm{angle}_{i,j,k}$ is the bending energy of FG-Nups, and $U^\mathrm{steric}_{i,j}$ is the steric repulsion.
The summations for the first and second terms are run over the neighboring beads pair within FG-Nups, while the summation for the third term are run over all beads pair in the system. 
Details of each potential is shown below:
\begin{eqnarray}
    U^\mathrm{bond}_{i,j} &=& k_b (r_{ij}-r_\mathrm{eq})^2\\
    U^\mathrm{angle}_{i,j,k} &=& k_a \qty{1-\cos (\theta_{ijk} - \theta_\mathrm{eq})}\\  
    U^\mathrm{steric}_{i,j} &=& 
    \begin{cases}
    U^\mathrm{rep}(r_{ij}), & \text{for $r_{ij}<r_\mathrm{cri}$}\\
    U^\mathrm{rep}(r_\mathrm{cri}), & \text{otherwise}
    \end{cases}\\
    U^\mathrm{rep}(r_{ij}) &=& \epsilon_r \exp (-r_{ij}/\sigma_r)
\end{eqnarray}
where $k_b$ is the spring constant, $r_{ij}$ is the distance between bead $i$ and $j$, $r_\mathrm{eq}$ is the equilibrium distance between two neighboring beads, $k_a$ is the bending rigidity, $\theta_{ijk}$ is the angle formed between bead $i$, $j$, and $k$, $\theta_\mathrm{eq}$ is the equilibrium angle, $\epsilon_r$ is the repulsion energy, $\sigma_r$ is the characteristic length for repulsion, and $r_\mathrm{cri}$ is the cutoff distance.

\begin{table*}[ht]
\centering
\begin{tabular}{cccc}
\hline\hline
Symbol & Value & Unit & Description\\
\hline
$L_x$ & 100 & nm & X-length of the simulation box\\
$L_y$ & 100 & nm & Y-length of the simulation box\\
$L_z$ & 160 & nm & Z-length of the simulation box\\
$h$ & 40 & nm & Thickness of the NE\\
$n_\mathrm{FG}$ & 80 & - & Number of FG-Nups\\
$l_\mathrm{segm}$ & 0.86 & nm & Diameter of the bead\\
$\zeta$ & 4.05$\times 10^{-17}$ & g/ns & Friction coefficient\\
$\kBT$ & 4.28$\times 10^{-18}$ & $\text{nm}^2$ g/$\text{ns}^2$ & Thermal energy at temperature $T$ = 310.15 (K)\\
$k_b$ & 100 & $\kBT/l_\mathrm{segm}^2$ & Spring constant of FG-Nups\\
$r_\mathrm{eq}$ & 1 & $l_\mathrm{segm}$ & Equilibrium distance between beads\\
$k_a$ & 0.5 & $\kBT$ & Bending rigidity of FG-Nups\\
$\theta_\mathrm{eq}$ & $\pi$ & rad & Equilibrium angle\\
$\epsilon_r$ & 400 & $\kBT$ & Repulsion energy\\
$\sigma_r$ & 1 & $l_\mathrm{segm}$ & Characteristic length for repulsion\\
$r_\mathrm{cri}$ & 1 & $l_\mathrm{segm}$ & Cutoff distance for repulsion\\
$r_\mathrm{thre}$ & 2 & $l_\mathrm{segm}$ & Threshold distance for FO $\leftrightarrow$ FI transition\\
$F_\mathrm{FG}$ & 10 & pN & Constant force between bound (B) stickers\\
$k_\mathrm{on}$ & $1.0 \times 10^{-1}$ & /ns & Rate constant for IF $\rightarrow$ B transition\\
$k_\mathrm{off}$ & $1.0 \times 10^{-2}$ & /ns & Rate constant for B $\rightarrow$ IF transition\\
$n_\mathrm{FFS}$ & 8 & - & Number of FFS interfaces\\
$t_\mathrm{basin}$ & $1.0 \times 10^5$ & ns & Duration for the basin simulation\\
$n_\mathrm{basin}$ & 256 & - & Number of trials for the basin simulation\\
$n_\mathrm{trials}$ & 50 & - & Number of trials for the probability calculation\\
$n_\mathrm{mesh}$ & 1000 & - & Number of random points for mesh calculation\\
\hline\hline
\end{tabular}
\caption{\label{tab:para} Parameters employed in our simulations. The values listed here were consistently used throughout our calculations.}
\end{table*}

\begin{table*}[ht]
\centering
\begin{tabular}{cccccccccc}
\hline\hline
Interface number $i$ & 0 & 1 & 2 & 3 & 4 & 5 & 6 & 7 & 8 \\
Number of trials $k_i$ & 50 & 20 & 20 & 20 & 20 & 15 & 15 & 15 & 10\\
\hline\hline
\end{tabular}
\caption{\label{tab:ffs}Parameters used for the FFS method.}
\end{table*}

\subsection*{Kinetic Monte Carlo}
In parallel with solving the Langevin equation, we employed the kinetic Monte Carlo method to simulate the dynamic interactions between stickers \cite{sgouros2019multiscale}.
Each sticker was categorized into one of three states: free outside (FO), free inside (FI), or bound (B). A bead (FO) was defined as a sticker with no other stickers within a distance of $r_\mathrm{thre}$ from its center, while a bead (FI) was defined as a sticker with more than one stickers within that distance. The FO $\leftrightarrow$ FI transition occurred via the diffusion described by Eq. \ref{eq:langevin}. Once a sticker became bound (B), a constant force $F_\mathrm{FG}$ towards the other attracted bead was applied. The FI $\leftrightarrow$ B transition occurred in a Monte Carlo fashion, with on and off rates of $k_\mathrm{on}$ and $k_\mathrm{off}$, respectively. We imposed a restriction that each sticker could only bind to one other sticker and that FO $\leftrightarrow$ B transitions were forbidden.

\subsection*{Forward flux sampling}
To calculate the transport rate of the cargo through the NPC, we employed Forward Flux Sampling (FFS) \cite{hussain2020studying,allen2009forward}, an efficient method for determining the rate constant for barrier-overcoming events.
In this method, an order parameter is set, which increases in value as the system progresses over the barrier.
We then define ``interfaces'' corresponding to specific ranges of the order parameter.
The simulation is run to determine the probability of the system transversing each of these interfaces.
The rate constant is then calculated as the product of the flux at which the system enters the first interface and the subsequent probabilities of transversing each interface.

In our system, we used the position of the transported cargo as the order parameter and defined the interfaces by the spatial regions in our simulation box.
We segmented the simulation space into several distinct regions as shown in Fig. \ref{fig:sim}B.
The 0-th interface region, denoted as $\mathcal{I}_0$, was designated as a half hemisphere with a radius of $D/2$ nm, originating from the entrance of the pore.
The area outside this hemisphere was set as the first basin, denoted as $\mathcal{B}$.
Subsequently, we divided the central pore region horizontally into eight equally sized segments, each shaped as a cylinder with a radius of $D/2$ nm and a height of $h/n_\mathrm{FFS}$ nm.
These cylindrical segments, arranged from the top to the bottom of the pore, were defined as the regions for the 1-st to $n_\mathrm{FFG}$-th interfaces, labeled as $\mathcal{I}_i$ ($i=1...n_\mathrm{FFS}$).
The region below the $n_\mathrm{FFS}$-th interface, or the space outside of the central pore region, was designated as the second basin.

For the FFS simulation, we initiated the process by simulating the system dynamics within the first basin, $\mathcal{B}$, over a period of $t_\mathrm{basin}$ seconds.
Every time the system crossed into the 0-th interface region, $\mathcal{I}_0$, we logged the system's coordinates, which includes the positions of both the cargo and the polymers.
This process was repeated $n_\mathrm{basin}$ times, each with a different simulation seed, to calculate the flux into the first interface, represented as $\Phi_0$.
We defined this flux, $\Phi_0$, as the total number of times the system entered $\mathcal{I}_0$, divided by the total simulation time, which is the product of $n_\mathrm{basin}$ and $t_\mathrm{basin}$.

We proceeded to calculate the probability that the system traverses from the 0-th interface to the 1-st interface, represented as $P(\lambda_1|\lambda_0)$.
Among the system's coordinates recorded previously, we randomly selected $n_\mathrm{trials}$ coordinates.
Each of these coordinates was designated as the starting point for $k_0$ independent simulations.
We then computed their dynamics until the system either retreated back into the basin $\mathcal{B}$ or crossed over to the next interface $\mathcal{I}_{1}$.
Whenever the system entered the next interface, we recorded the system's coordinates.
The transverse probability, $P(\lambda_1|\lambda_0)$, was calculated as the ratio of successful trials (those that reached $\mathcal{I}_1$) to the total number of trials.
For the calculation of subsequent transverse probabilities, $P(\lambda_{i+1}|\lambda_i)$ ($i=1...n_\mathrm{FFS}$), we employed a similar process.
For each interface, we launched $k_i$ trials. 
When calculating the transverse probability, we factored the weight for each trial as defined by the Rosenbluth process \cite{allen2006simulating}.

Finally, we computed the transport rate constant, $k_{AB}$, as the product of the initial flux into the 0-th interface and the cumulative transverse probability across all interfaces. 
This is represented by the equation:
\begin{equation}
k_{AB} = \Phi_0 \prod_{i=0}^{n_\mathrm{FFS}} P(\lambda_{i+1}|\lambda_i).
\end{equation}
However, it's worth noting that, after the section of the results subtitled ``Transport rate of inert cargoes", we modified our definition of the transport rate to $k_{AB} = \Phi_0 P(\lambda_{1}|\lambda_0)$, where transverse probabilities for higher orders are omitted. 
This modification allowed us to focus on the translocation across the critical entrance region, which we believed is the major factor determining the overall transport rate.

\subsection*{Mesh size measurement}
To quantitatively evaluate the mesh size within our simulation system, we employed the following method, which is based on the approach proposed by Sorichetti et al. \cite{sorichetti2020determining}.
Initially, we prepared a coarse-grained NPC pore with equilibrated FG-Nups. 
A random point $\rr_p$ in the central channel was then selected. 
If this selected point overlapped with the beads constituting FG-Nups, it was discarded, and a new point was chosen. 
This process was repeated until a point with no overlap was found.
Subsequently, we identified the largest sphere, with radius $r(\rr_c)$, that contained the chosen point $\rr_p$ but did not overlap with any system beads. 
This process can be formally expressed as maximizing the function below:
\begin{equation}
r(\rr_c) = \min {\abs{\rr_c - \rr_i}} - l_\mathrm{segm}/2,
\end{equation}
subject to the constraint,
\begin{equation}
\abs{\rr_c - \rr_p} - r(\rr_c) \le 0.
\end{equation}
In the equations above, $\rr_c$ represents the center of the sphere. 
We treated this as a nonlinear optimization problem and solved it using the Sbplx algorithm from the NLopt library \cite{NLopt,SUBPLEX}.
The procedure was repeated for $n_\mathrm{mesh}$ times, with different chosen positions $\rr_p$, and we computed the average of $r(\rr_c)$. 
This average, when multiplied by two, was then defined as the mesh size $\xi$.

\section{Results\label{sec:res}}

\subsection*{Polymer characteristics}
Firstly, we studied the scaling behavior of the polymers.
We prepared polymers of different sizes ($N$ = 10, 18, 30, 54, 100, 180, 300).
Each polymer was placed in a bulk space (without the NPC boundary), where the dynamics and structure of individual polymers were simulated.
The simulation procedure was as follows:
each polymer was initially placed in its fully extended state.
The system was then equilibrated for 50 \si{\micro\second} to allow for relaxation of the polymer structure.
Following this, we simulated the system for an additional 100 \si{\micro\second}, and sampled the polymer conformation every 1 \si{\micro\second}.
For each sampled conformation, we calculated the radius of gyration and then took the average across all the sampled conformations.

\begin{figure}[t]
\includegraphics[width=0.35\textwidth]{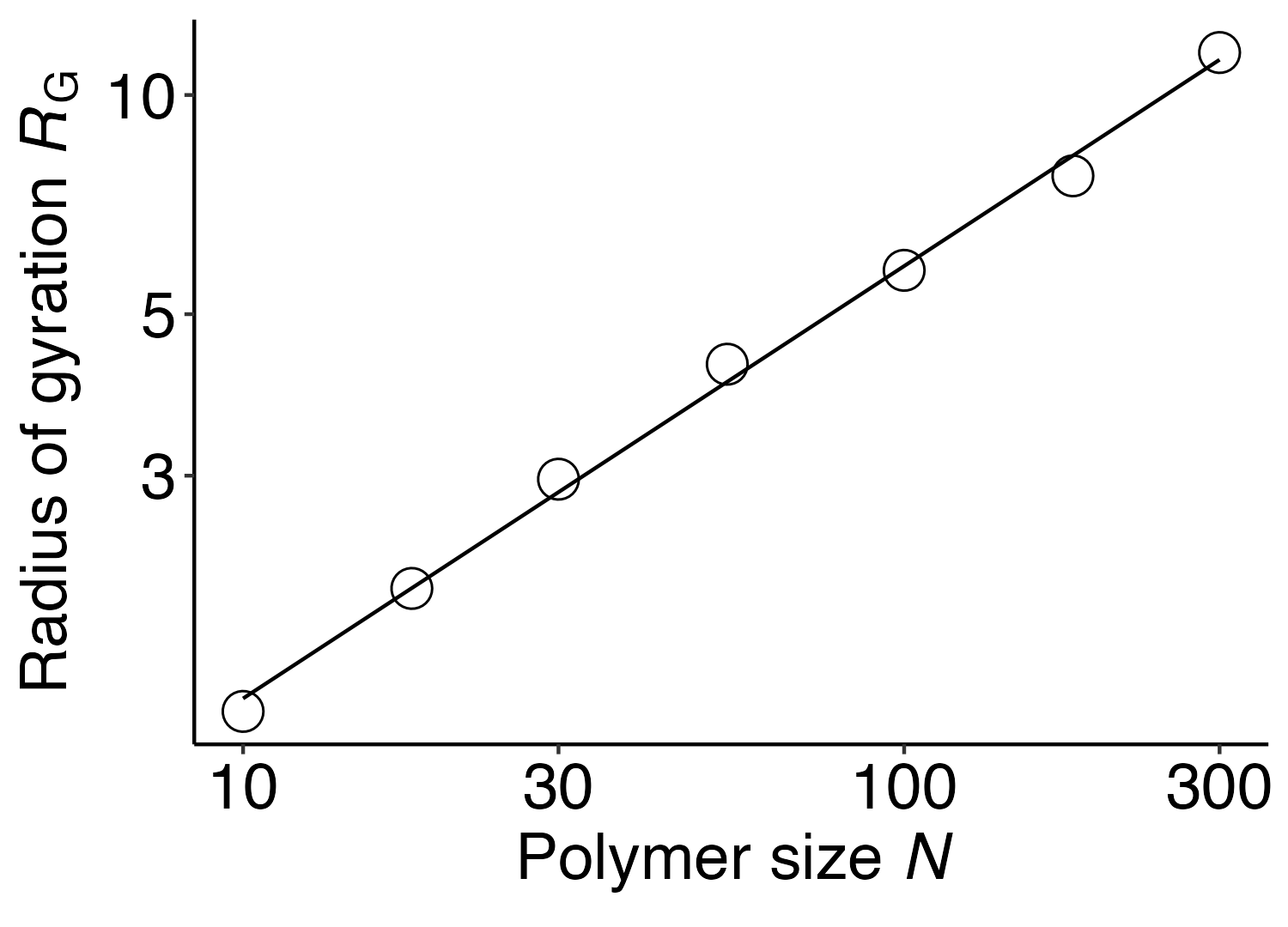}
\caption{Scaling relationship between the polymer size $N$ and the radius of gyration $R_\mathrm{G}$. Data is presented on a logarithmic scale. We applied linear regression to fit the data, yielding a slope of 0.5939, which closely aligns with the widely accepted Flory exponent of $\nu \cong$ 0.588.}
\label{fig:scale}
\end{figure}

The relationship between polymer size and the average radius of gyration is depicted in Fig. \ref{fig:scale}.
In the log-log plot, the data points align linearly, displaying the expected scaling behavior of the polymers, $R\sim N^\nu$.
By fitting the data with a straight line using the least squares method, we determined the slope of the fitted line to be 0.5939.
As this value is quite close to the commonly used Flory exponent for swollen linear chains \cite{rubinstein2003polymer}, which is 0.588, we assume $\nu \cong$ 0.588 in the ensuing discussion.

\begin{figure*}[t]
\includegraphics[width=0.9\textwidth]{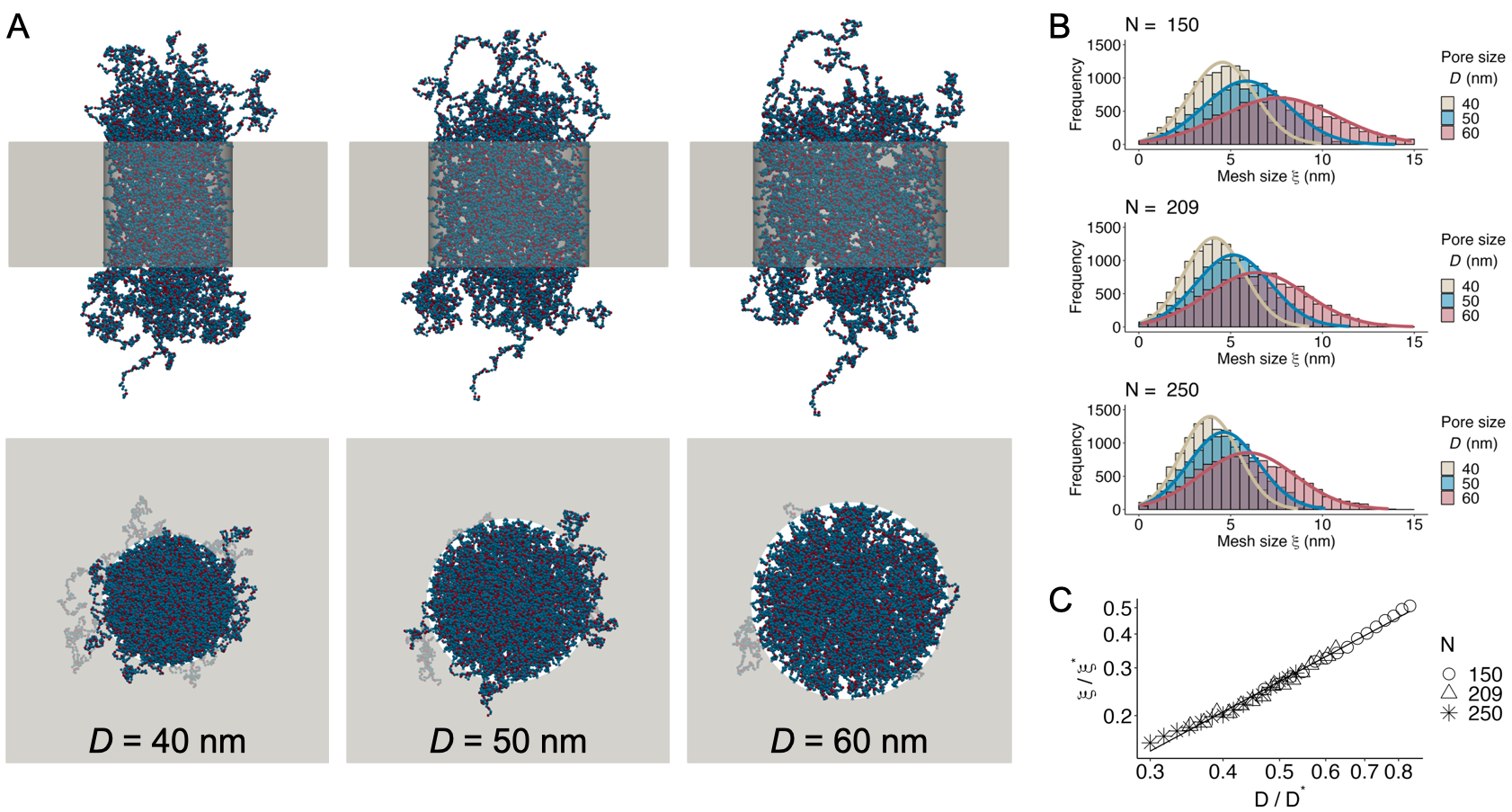}
\caption{Measurement and analysis of mesh size. (A) FG-Nup conformations within the NPC at varying diameters, shown from front and top perspectives. (B) Distribution of mesh sizes corresponding to each pore size. The histogram represents calculated data, while the solid line denotes the Gaussian fitting model. (C) Relationship between normalized pore size $D$ and mesh size $\xi$, both normalized by reference lengths $D^*$ and $\xi^*$ respectively. Data is presented on a logarithmic scale.}
\label{fig:mesh}
\end{figure*}

\subsection*{Mesh size}
Next, we examined the relationship between the mesh size and the pore size.
We constructed the NPC system with varying polymer lengths ($N$ = 150, 209, and 250) and a diverse set of pore diameters ($D$ = 36, 38, 40, 42, 44, 46, 48, 50, 52, 54, 56, 58, 60, 62, and 64 nm).
The selection of these parameters was informed by existing biological data:
studies on yeast NPCs reveal that the average and standard deviation of FG-Nup length are approximately 180 and 10 nm, respectively \cite{yamada2010bimodal}.
These measurements correspond to $N$ values of 209 and 11.
We chose $N$ = 150, 209, and 250 to adequately cover this range. 
As for the pore diameter, typical measurements fall within the range of $D$ = 40-60 nm \cite{matsuda2022nuclear}. 
This range is effectively covered by the parameters selected for our study.
We ensured that all the pore diameters examined in this study were smaller than the pore diameter at overlap concentration, $D^*$. 
For $N$ = 150, 209, and 250, the pore diameter at overlap concentration corresponds to $D^*$ = 76.4, 102, and 120 nm, respectively, which is larger than all pore diameters examined.

For each NPC configuration, we allowed the system to equilibrate for 50 \us{} before conducting the production run for 100 \us.
Position of the all beads were recorded every 10 \us, and we measured the mesh size for each of these configurations.
The method employed to measure the mesh size is detailed in Section \ref{sec:sim} (Mesh size measurement).

We present the distribution of the sampled mesh sizes in Fig. \ref{fig:mesh}B, showcasing specific instances for $D$ = 40, 50, and 60 nm.
In all cases, the distribution exhibited a single peak, which could be effectively fitted using a Gaussian distribution.
Following this, we computed the mean of the fitted Gaussian distribution, which we defined as the mesh size $\xi$.
The relationship between the pore size $D$ and the mesh size $\xi$ is depicted in Fig. \ref{fig:mesh}. In Fig. \ref{fig:mesh}, we display the normalized values using the reference lengths, $D^* = A^{1/2}N^{3\nu/2}$ and $\xi^* = b N^\nu$.
In the logarithmic scale plot, $\xi/\xi^*$ and $D/D^*$ exhibit a linear relationship, lending support to our model (Eq. \ref{eq:xi}). 
The slope of the fitted line was calculated to be 1.158, which is in reasonable agreement with our model prediction, $2\nu/(3\nu-1) \cong 1.539$.



\subsection*{Transverse probability}
Next, we set out to validate our hypothesis that the insertion of the cargo into the polymer solution is the main determinant of the transport rate. 
To investigate this, we used a forward flux sampling (FFS) simulation \cite{hussain2020studying,allen2009forward} to compute the probability of the cargo traversing each sub-region within the NPC. 
Through comparisons of the transverse probabilities for each interface, we aimed to pinpoint the process that contributes most significantly to the variability in the cargo's transport through the NPC.

\begin{figure}[t]
\centering
\includegraphics[width=0.48\textwidth]{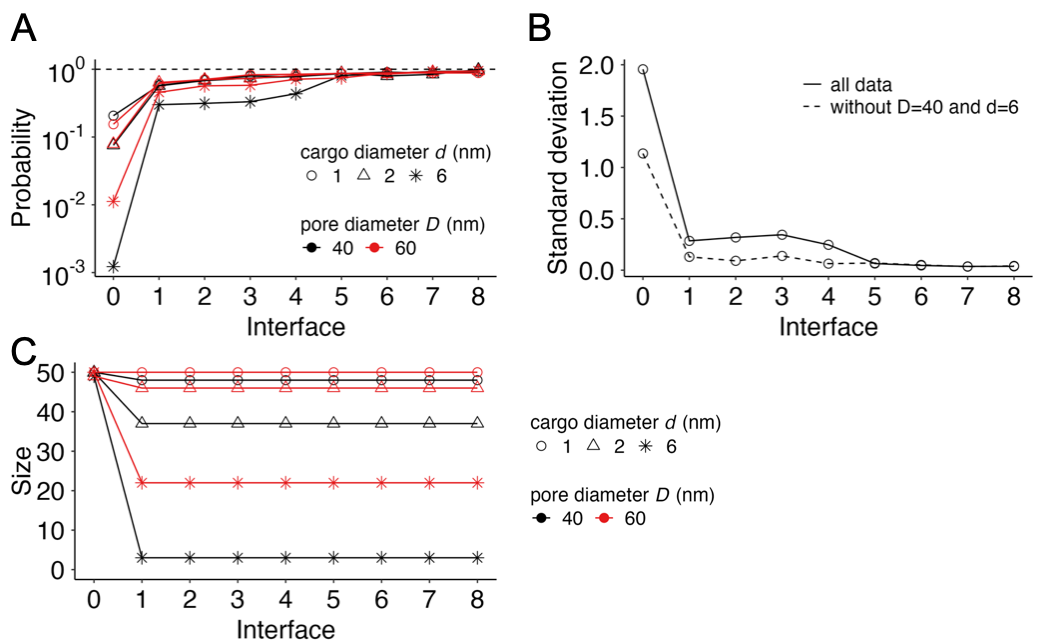}
\caption{Analysis of transverse probability during cargo transport through the NPC. (A) Transverse probability, $P(\lambda_{i+1}|\lambda_i)$, for each interface, $i$ = 0, 1, ..., 8, with a polymer length of $N$ = 150. The dashed line indicates a probability of 1. (B) Standard deviation of $\ln P(\lambda_{i+1}|\lambda_i)$ for each interface $i$ across different parameter sets. The solid line represents all data, while the dashed line excludes the case $(D,d) = (40, 6)$. (C) Sample size used in the calculation of transverse probability.}
\label{fig:prog}
\end{figure}

We performed the FFS simulation for various cargo sizes, $d$ = 1, 2, 6 nm, and pore sizes, $D$ = 40, 60 nm.
We set the polymer length to be $N$ = 150. 
We observed that the transverse probability, $P(\lambda_{i+1}|\lambda_i)$, exhibited the greatest variability across different parameters when the cargo was navigating through the 0-th interface. 
In contrast, the transverse probabilities were relatively similar across different parameters for the other interfaces (Fig. \ref{fig:prog}A).
This trend can be more clearly visualized by considering the standard deviation in the logarithm of the transverse probability for each interface, $\ln P(\lambda_{i+1}|\lambda_i)$, as illustrated in Fig. \ref{fig:prog}B.
However, we noted that for the parameter set $(d, D)$ = (6, 40), the sample size for the simulation of $\mathcal{I}_i$ ($i\ge1$) was quite limited (Fig. \ref{fig:prog}C). 
This could potentially distort the calculated transverse probability, leading it to deviate from the expected behavior. 
To mitigate this, we excluded this case and recalculated the standard deviation (Fig. \ref{fig:prog}B), which more unequivocally demonstrated that the variability in transverse probability was primarily attributed to the passage through the 0-th interface.

This finding suggests that the primary determinant of the overall transport rate is the initial insertion of the cargo into the central channel, while the transverse processes within the central channel occur with equal probability among the different parameter sets.
This is in line with our original assumptions made during the model formulation (Eq. \ref{eq:rate_ori}).

Given this observation, for the purposes of our discussion in the following sections, we choose to redefine the transport rate as $k_\mathrm{AB} = \Phi_0 P(\lambda_{1}|\lambda_0)$. In this reformulation, we consciously omit the transverse probabilities for $i\ge1$, focusing solely on the initial entrance of the cargo into the central channel.

\begin{figure*}[ht]
\includegraphics[width=0.9\textwidth]{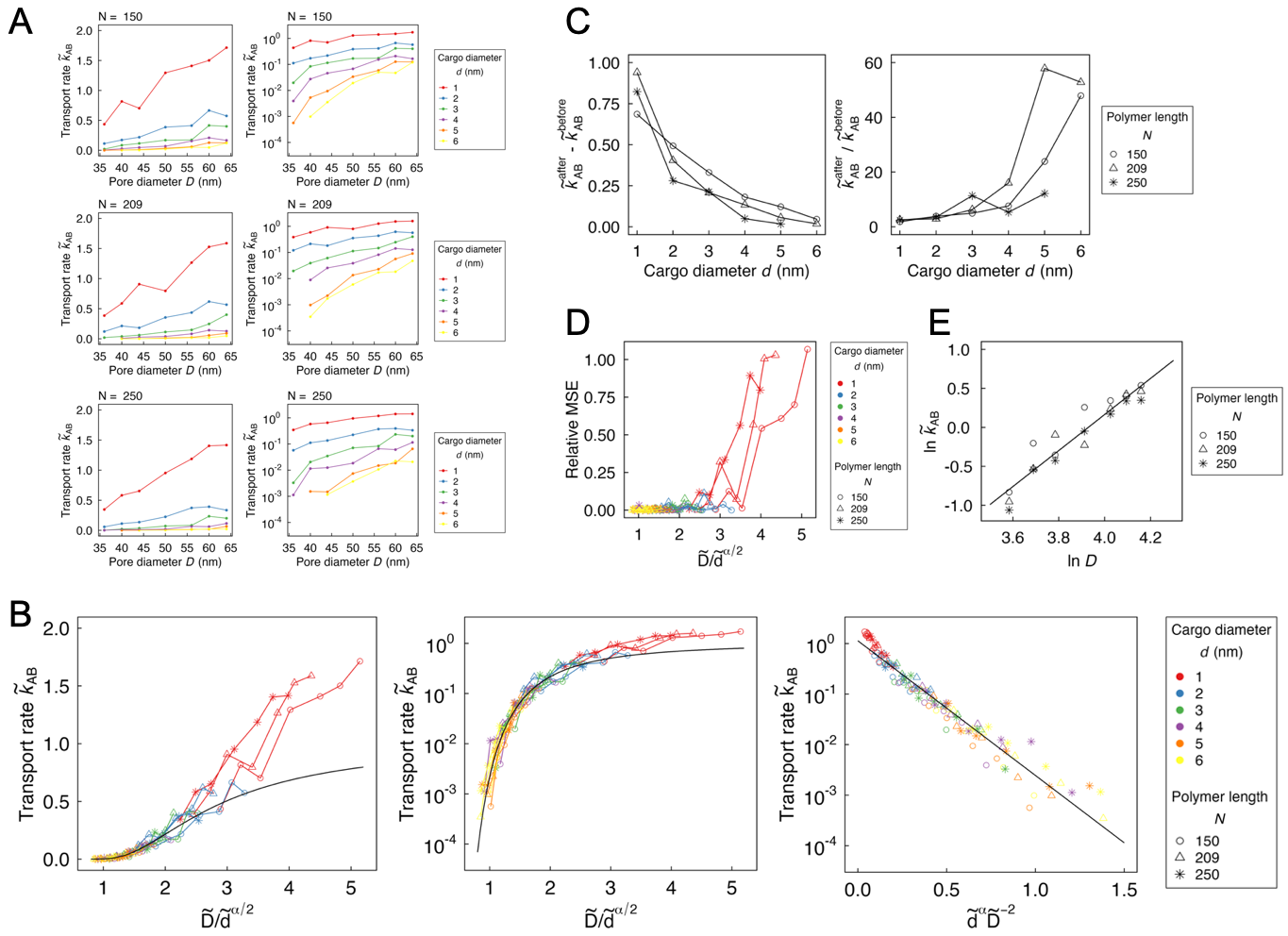}
\caption{Transport rate analysis for the non-interacting cargo. (A) Transport rate as a function of the pore diameter with vertical axis represented in both linear (left) and logarithmic scales (right). (B) Transport rate depicted as a function of $\tilde{D}/\tilde{d}^{\alpha/2}$ in linear scale (left), $\tilde{D}/\tilde{d}^{\alpha/2}$ in semi-logarithmic scale (middle), and $\tilde{d}^{\alpha}\tilde{D}^{-2}$ in semi-logarithmic scale (right). Black solid line corresponds to the model formula (Eq. \ref{eq:k_AB}) with $\gamma$ = 6.13. (C) Change in the transport rate from $D=40$ nm ($\tilde{k}_\mathrm{AB}^\mathrm{before}$) to $D=60$ nm ($\tilde{k}_\mathrm{AB}^\mathrm{before}$), with left figure showing the absolute difference and right figure showing the increment rate. (D) Relative mean square error as a function of $\tilde{D}/\tilde{d}^{\alpha/2}$. The error is calculated as the difference between each data point and the model formula with $\gamma$ set to 6.13. Vertical axis represents the mean square error divided by the value predicted by the model. (E) Transport rate versus pore size for the case of $d$ = 1 nm, represented in logarithmic scale. The solid line corresponds to the linear regression of the data, yielding a slope of 2.31.}
\label{fig:rate}
\end{figure*}

\subsection*{Transport rate of non-interacting cargoes}
Next, we explored the relationship between the transport rate and the pore size.    
In this section, we studied the transport rate of inert cargo, meaning that there is no attractive interaction between the cargo and FG-Nups. 
The systems under examination were parameterized with the following sets: $N$ = 150, 209, 250, $D$ = 36, 40, 44, 50, 56, 60, 64 nm, and $d$ = 1, 2, 3, 4, 5 nm.
The simulation procedure was as follows: We initially set up the system without the transported molecule and allowed the FG-Nup conformation to equilibrate for 50 \us. We then placed the transported molecule at a fixed position, 50 nm away from the pore opening. The system was then allowed to equilibrate for 1 \us{} with the position of the transported molecule held fixed.
Following this equilibration process, we conducted the Forward Flux Sampling (FFS) simulation \cite{hussain2020studying,allen2009forward}.
As previously observed, the transverse probability tends to converge to a certain value after the 1-st interface. 
Therefore, from this point forward, we re-defined the transport rate as the product of the flux and the probability to transverse the first interface, $k_\mathrm{AB} = \Phi_0 \times P(\lambda_1|\lambda_0)$.

The computed transport rates are displayed in Fig. \ref{fig:rate}A. 
Here, we represent the normalized transport rate, symbolized as  $\tilde{k}_\mathrm{AB} = k_\mathrm{AB}/k_0$.
The normalization factor, $k_0$, is set at $10^{-5}$ (1/ns), which serves as our baseline for comparison.
Our findings show that there is an increase in the transport rate with the enlargement of the pore diameter. 
Interestingly, we noticed that the magnitude of this increase is contingent on the size of the cargo being transported.
When evaluating the data in a linear scale, we find that the absolute increment in the transport rate, given by $\tilde{k}_\mathrm{AB}^\mathrm{after} - \tilde{k}_\mathrm{AB}^\mathrm{before}$, is more pronounced for smaller cargoes. 
On the contrary, when we examine the data on a logarithmic scale, the relative increment rate, calculated as $\tilde{k}_\mathrm{AB}^\mathrm{after} / \tilde{k}_\mathrm{AB}^\mathrm{before}$, is more prominent for larger cargoes (Fig. \ref{fig:rate} C). 
This suggests that while smaller cargoes gain more in absolute terms with an increase in pore diameter, larger cargoes enjoy a greater relative improvement.

We subsequently examined the accuracy of our model equation (Eq. \ref{eq:k_AB}) by graphing the transport rate in relation to $d^\alpha D^{-2}$ (Fig. \ref{fig:rate}B, right panel).
When plotted on a semi-logarithmic scale, the transport rate, $\tilde{k}_\mathrm{AB}$, displayed a linear correlation with $d^\alpha D^{-2}$. 
This finding demonstrates that $d^\alpha D^{-2}$ is indeed exponentially related to $\tilde{k}_\mathrm{AB}$, as predicted by Eq. \ref{eq:k_AB}.
Importantly, data points corresponding to various polymer lengths $N$ all aligned along a single line, which attests to the universal applicability of Eq. \ref{eq:k_AB}.
Utilizing the least square method, we fitted the data to a line and calculated its slope, which yielded a $\gamma$ value of 6.13.

While our data generally adhered to the model equation (indicated as the black line in Fig. \ref{fig:rate}B), there was a significant deviation observed, especially for smaller cargoes with $d$ = 1 nm.
This discrepancy was particularly noticeable when evaluating the relative mean square error (Fig. \ref{fig:rate}D).
For this calculation, we measured the error between our empirical data and the predictions made by our model formula (Eq. \ref{eq:k_AB}). 
With $\gamma$ set to 6.13, we computed the mean square of these errors and normalized this value by the prediction given by the model. 

In our Discussion section (\ref{sec:disc}), we speculated that this divergence can be attributed to the notion that for relatively small cargoes, the structure of the polymer network does not function as an effective barrier impeding their transit through the NPC.
Consequently, the free energy required for cargo insertion (Eq. \ref{eq:free}) is lessened, causing the transport rate to deviate from our model equation (Eq. \ref{eq:k_AB}).
Under these circumstances, it's plausible that the global structure of the pore, namely the pore diameter $D$, could directly influence the transport rate, rather than the local mesh structure $\xi$.
When the size of the cargo is significantly smaller than the pore size, we anticipate that the transport rate is proportional to the entrance area of the pore, that is, $k_\mathrm{AB}\sim D^2$.
To validate this assumption, we plotted $\ln \tilde{k}_\mathrm{AB}$ against $\ln D$ specifically for the instance where $d$ = 1 nm (Fig. \ref{fig:rate}E).
By fitting the data to a line using the least square method, we determined its slope to be 2.31.
This value is close enough to our prediction of 2, reinforcing the validity of our hypothesis.

\subsection*{Transport rate of interacting cargoes}
Finally, we calculated the transport rate for cargoes demonstrating attractive interactions with FG-motifs. 
We introduced an attractive surface area, $S$, on the cargo and designated the beads corresponding to this attractive surface as ``attractive beads''. 
These attractive beads interact with FG-Nups in the same manner as the stickers in FG-Nups.
Simulations were carried out with FG-Nup lengths of $N$ = 209, pore sizes $D$ ranging from 36 nm to 64 nm, cargo sizes $d$ from 2 nm to 6 nm, and the attractive surface area $S/S^*$ = 1, where $S^*$ = 4$\pi$ $\mathrm{nm}^2$ represents the surface area of the cargo with $d$ = 2 nm.
As in previous calculations, we computed the transverse probability up to the first interface and determined the transport rate, $k_\mathrm{AB}$, as the product of $\Phi_0$ and $P(\lambda_1|\lambda_0)$.

\begin{figure}[t]
\centering
\includegraphics[width=0.49\textwidth]{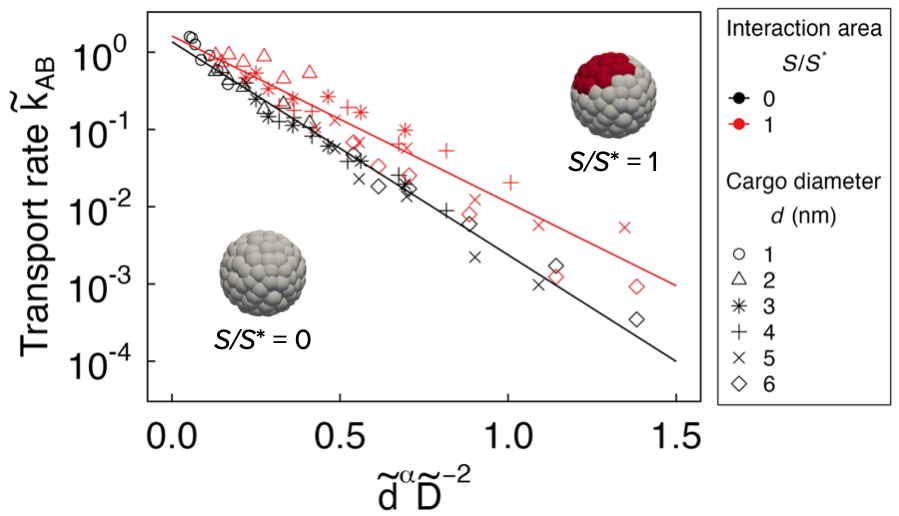}
\caption{Comparison of transport rates for interacting and non-interacting cargoes. 
FG-Nup length is $N$ = 209.
The solid line represents linear regression of the data, yielding a slope of -6.35 for non-interacting cargoes (black) and -4.96 for interacting cargoes (red). Inset images depict the coarse-grained surface structure of the cargo with $d$ = 5 nm, highlighting grey and red beads representing non-interacting and interacting surface regions, respectively.
}
\label{fig:active}
\end{figure}

The results, including those for non-interacting cargoes, are presented in Fig. \ref{fig:active}. 
The transport rate for interacting cargoes exhibited an exponential dependence on $\tilde{d}^\alpha \tilde{D}^{-2}$, indicating that Eq. \ref{eq:k_AB} is also applicable to interacting cargoes within the simulated parameter range. 
The fitting line's slope, i.e. parameter $\gamma$, was calculated to be 4.96.
This value is smaller than the $\gamma$ value for non-interacting cargoes (computed solely for $N$ = 209), which stands at 6.35. 
This observation aligns with our intuitive understanding that the presence of attractive interactions between the cargo and FG-Nups lowers the free energy required for cargo insertion into the pore, resulting in a smaller $\gamma$ value.

\section{Discussion\label{sec:disc}}
\begin{figure}[t]
\centering
\includegraphics[width=0.49\textwidth]{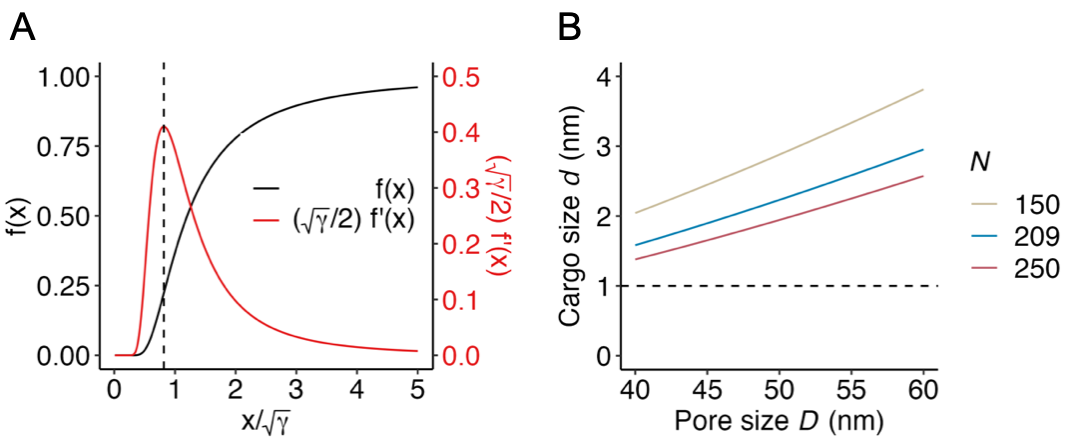}
\caption{Influence of pore dilation for different cargo sizes. (A) The model formula, $\tilde{k}_\mathrm{AB} = f(x)$, plotted as a function of $x=\tilde{D}/\tilde{d}^{\alpha/2}$ (black) and its derivative (red). The dashed line represents $x=x_0 =\sqrt{2\gamma/3}$, which gives the peak of $f'(x)$. (B) Cargo size that corresponds to $x=x_0$ for each pore size. The dashed line represents $d=1$ nm.}
\label{fig:math}
\end{figure}

Our results suggest that 1) the extent of the change in the transport rate is contingent on the size of the cargo and 2) for smaller cargoes ($d$ = 1 nm), the simulation results diverge from our theoretical model.
To delve deeper into these aspects, we can rewrite Eq. \ref{eq:k_AB} as:
\begin{equation}
\label{eq:math}
\tilde{k}_\mathrm{AB} = 
f(x) = 
\exp (-\gamma/x^2).
\end{equation}
Here, $x\equiv \tilde{D}/\tilde{d}^{\alpha/2}$ represents the pore diameter, adequately renormalized by the cargo size.
This equation and its derivative are depicted in Fig. \ref{fig:math}A.
The transport rate function, $f(x)$, grows monotonically with $x$, and its derivative, $f'(x)$, exhibits a pronounced peak at $x_0=\sqrt{2\gamma/3}$. 
This pattern suggests that while pore dilation consistently increases the transport rate, its effect is most noticeable at a certain cargo size, i.e. when $x$ approximates $x_0$.
For significantly larger or smaller cargoes, $x$ deviates substantially from $x_0$, leading to a less pronounced change in the transport rate. 
This interpretation is consistent with intuitive reasoning: significantly large cargoes are blocked from entering the pore irrespective of its expansion. Conversely, considerably small cargoes, which are always permeable through the NPC, see little to no alteration in their transport rate due to changes in the local mesh structure.
It should be noted that since the derivative of the logarithm of the transport rate, $(\ln f(x))'= 2\gamma / x$, decreases monotonically with $x$, 
the relative increment rate, $\tilde{k}_\mathrm{AB}^\mathrm{after} / \tilde{k}_\mathrm{AB}^\mathrm{before}$, increases with the cargo size as shown in our simulation result. 

Although the derivative $f'(x)$ peaks at $x=x_0$, our results showed a consistent increase in $\tilde{k}_\mathrm{AB}^\mathrm{after} - \tilde{k}_\mathrm{AB}^\mathrm{before}$. 
We attribute this phenomenon to alterations in the global structure of the pore. 
When $x$ exceeds $x_0$, the impact of the mesh size diminishes, which is signified by the decreasing trend of $f'(x)$.
In such circumstances, the impact of the global structure, which we omitted during our model formulation, starts to gain significance. 
In this regime, it is reasonable to hypothesize that the transport rate scales with the area of the pore entrance, i.e., $\tilde{k}_\mathrm{AB} \sim D^2$, instead of Eq. \ref{eq:k_AB}.
This proposition was supported by our simulation results, as seen in Fig. \ref{fig:rate}E. 
To reinforce this hypothesis, we computed the cargo size $d$ corresponding to $x=x_0$ for each pore size $D$ (Fig. \ref{fig:math}B). 
For all polymer length $N$, $d$ = 1 nm was was found to be smaller than this critical value. 
This observation aligns with our simulation results, where the most pronounced deviation from the model formula was observed for $d$ = 1 nm.

\section{Conclusion\label{sec:conc}}
In this study, we proposed a model equation (Eq. \ref{eq:k_AB}) that elucidates the influence of pore dilation on the transport rate. This formula is particularly relevant when the pore size $D$ is less than the critical diameter $D^*$, below which the individual FG-Nups within the central channel come into contact with each other.
The range where this model applies aligns with the experimentally observed dilation range of the NPC, leading us to believe that our model can be employed for the analysis of the natural NPC.

The model equation was derived using polymer scaling theory \cite{rubinstein2003polymer,de1979scaling,doi1988theory}.
Its formulation primarily rests on two key assumptions: (1) the mesh size $\xi$ scales with the pore size $D$ in a specific manner, as shown in Eq. \ref{eq:xi}, and (2) the primary free energy barrier for cargo passing through the NPC arises when the cargo enters the pore. We validated each of these assumptions, as well as the final model equation (Eq. \ref{eq:k_AB}), through computer simulations.

The limitation of our model is that it fails to account for significantly small cargoes, for which the mesh size does not act as a barrier, leading to transport rates that exceed the model's predictions. 
We can roughly estimate if the cargo of interest is significantly small by computing the parameter $x=\tilde{D}/\tilde{d}^{\alpha/2}$, and determining if it is less than the critical value $x_0 = \sqrt{2\gamma/3}$.

Looking forward, future research should aim to validate and refine this model experimentally, with the goal of consolidating its utility and addressing its current limitations. By continuing along this line of inquiry, we aspire to deepen our understanding of the intricate relationship between NPC dilation and molecular transport. 

\section*{Competing Interest}
The authors declare no competing interest.

\section*{Data, Materials, and Software Availability}
All study data are included in the article.

\section*{Acknowledgement}
This work used Expanse at San Diego Supercomputer Center through allocation MCB100146 from the Advanced Cyberinfrastructure Coordination Ecosystem: Services \& Support (ACCESS) program, which is supported by National Science Foundation grants 2138259, 2138286, 2138307, 2137603, and 2138296.

\bibliography{bib}
\end{document}